\documentclass{osa-article}
\usepackage{soul}
\journal{osajournal}
\setmojecopyright

\articletype{Research Article}
\usepackage{xcolor}
\usepackage{url}

\usepackage[normalem]{ulem}
\usepackage{caption}
\usepackage{subcaption}

\usepackage[final]{changes}  % highlight  changes
\definechangesauthor[color=blue, name={Marcin Witkowski}]{mw} 
\usepackage{todonotes}
\makeatletter
\setcommentmarkup{%
  \todo[
    textcolor={authorcolor!70!black},
    backgroundcolor={authorcolor!20!white},
    linecolor={authorcolor!30!white},
    size=\scriptsize]%
  {\@nameuse{Changes@AuthorName#2}\\#1}
}
\makeatother

\begin{document}
%\title%{\sout{Photoionisation losses in a strontium optical lattice clock operating at blue-detuned magic wavelength light} 
\title {{Photoionisation cross sections of ultracold $^{88}$Sr in $^1$P$_1$ and $^3$S$_1$ states at 390 nm and the resulting  blue-detuned magic wavelength optical lattice clock constraints}}

\author{Marcin Witkowski\authormark{1,*},  S\l{}awomir Bilicki\authormark{1}, Marcin Bober\authormark{1}, Domagoj Kova\u{c}i\'c\authormark{1,2}, Vijay Singh\authormark{1}, Ara Tonoyan\authormark{1,3}, and  Micha\l{} Zawada\authormark{1}}
%\author{M.~Witkowski\authormark{1,*},  S.~Bilicki\authormark{1}, M.~Bober\authormark{1}, D. Kova\u{c}i\'c\authormark{1,2}, V.~Singh\authormark{1}, A.~Tonoyan\authormark{1,3}, and  M.~Zawada\authormark{1}}

\address{\authormark{1}Institute of Physics, Faculty of Physics, Astronomy and Informatics, Nicolaus Copernicus University, Grudzi\c{a}dzka 5, PL-87-100 Toru\'n, Poland\\
\authormark{2}Institute of Physics, Bijeni\u{c}ka cesta 46, 10000 Zagreb, Croatia\\
\authormark{3}Institute for Physical Research, National Academy of Sciences of the Republic of Armenia, Ashtarak-2, 0203, Armenia\\
%\authormark{3}Institute for Physical Research, Armenian National Academy of Sciences, Ashtarak-2, 0203, Armenia\\
}

\email{\authormark{*}marcin\_w@umk.pl} %% email address is required

% \homepage{http:...} %% author's URL, if desired

%%%%%%%%%%%%%%%%%%% abstract %%%%%%%%%%%%%%%%
%% [use \begin{abstract*}...\end{abstract*} if exempt from copyright]

\begin{abstract}
    %\sout{We present the measurements of light-induced atomic losses of $^{88}$Sr at blue magic 389.889~nm wavelength due to photoionisation. The photoionisation cross sections for the excited states $^1$P$_1$ and $^3$S$_1$ are measured to be 2.20(50)$\times$10$^{-20}$~m$^2$ and 4.5(4.5)$\times$10$^{-23}$~m$^2$, respectively.
    %These values have been used to determine the impact of the blue magic-wavelength trapping light on the performance of the Sr optical lattice clocks for typical experimental conditions.}
    {We present the measurements of the photoionisation cross sections of 
    the excited $^1$P$_1$ and $^3$S$_1$ states of 
    ultracold $^{88}$Sr atoms at  389.889~nm wavelength, {which is the magic wavelength of the ${}^{1}$S${}_{0}$-${}^{3}$P${}_{0}$ clock transition}.
    The photoionisation cross section 
    of the $^1$P$_1$ state is determined from the measured ionisation rates 
    of $^{88}$Sr {in the magneto-optical trap}
    in the $^1$P$_1$ state to be 2.20(50)$\times$10$^{-20}$~m$^2$, while the photoionisation cross section of $^{88}$Sr in the $^3$S$_1$ state is  inferred from the photoionisation-induced reduction in the number of atoms transferred through the $^3\text{S}_1$ state %from the change in transition probability of the $^1$S$_0$-$^3$P$_0$ transition 
    {in an operating optical lattice clock}  to be $1.38(66)\times$10$^{-18}$~m$^2$.
    %4.5(4.5)$\times$10$^{-23}$~m$^2$.
     Furthermore, the {resulting limitations} of employing a blue-detuned magic wavelength optical lattice in strontium optical lattice clocks are evaluated. {We estimated photoionisation induced loss rates of atoms at 389.889~nm wavelength under typical experimental conditions 
     %to be $2.4\times$10$^{5}$~s$^{-1}$ for the $^1$P$_1$ and $1.26\times$10$^{8}$~s${^{-1}}$ for $^3$S$_1$ state 
     and made several suggestions on how to mitigate these losses. In particular, the large photoionisation induced losses for the $^3$S$_1$ state would make the use of the  $^3$S$_1$ state in the optical cycle in a blue-detuned optical lattice unfeasible and would instead require the less commonly used $^3$D$_{1,2}$ states during the detection part of the optical clock cycle. }} 
\end{abstract}

%%%%%%%%%%%%%%%%%%%%%%%%%%  body  %%%%%%%%%%%%%%%%%%%%%%%%%%
\section{Introduction}
Optical atomic clocks have proven to be excellent research instruments in various fields ranging from ultra-precise metrology~\cite{Nicholson2015,Schioppo2016,Lisdat2016,Grotti2018,Katori2020},
%\textcolor{green}%{\sout{through}} 
tests of fundamental physics~\cite{Delva2017,Katori2020} to searches for new matter in astrophysics~\cite{Wcislo2016,Wcislo2018, Roberts2020, Kennedy2020}. Along with improving atomic optical clocks, the factors limiting their performance have become  increasingly important. The perturbing effects induced by the environment, like light shifts~\cite{Shi2015}, 
%\textcolor{cyan}{
black-body radiation (BBR)
%} 
shifts~\cite{Middelmann2012}, and  collisional shifts~\cite{Gibble2013}, are matters of growing experimental concern. Several solutions have been implemented to counteract these undesirable effects, like Pauli blocking mechanism~\cite{Gibble1995,Gupta2003}, temperature decreasing through cryogenic systems~\cite{Ushijima2015}, and light shift cancellation by magic-wavelength-based optical traps~\cite{Katori2003,Ushijima2018}. 

A blue-detuned optical trap has been proposed as an alternative 
%\textcolor{green}{dipole trap in optical clocks} 
to commonly used red-detuned optical traps~\cite{Takamoto2009}. In blue-detuned optical traps, atoms are confined near the minimum of  light intensity, leading to a significant reduction of light-induced perturbations. On the other hand, if the energy of the 
%\textcolor{green}{photon} 
blue-detuned trapping light 
%\textcolor{green}{\sout{photon}} 
photon is above the atomic ionisation threshold, it may lead to atomic losses. 
Moreover, they were proposed together with the red-detuned conveyor belt optical lattice as a way to realise an active optical atomic clock based on  superradiance~\cite{Kazakov13}.
In the present work, we analyse the feasibility of the blue-detuned magic wavelength for the optical trapping
%\textcolor{green}{of} 
of strontium atoms and possible atomic losses due to photoionisation.

\begin{figure}[hbt]
 \centering\includegraphics[width=0.6\columnwidth]{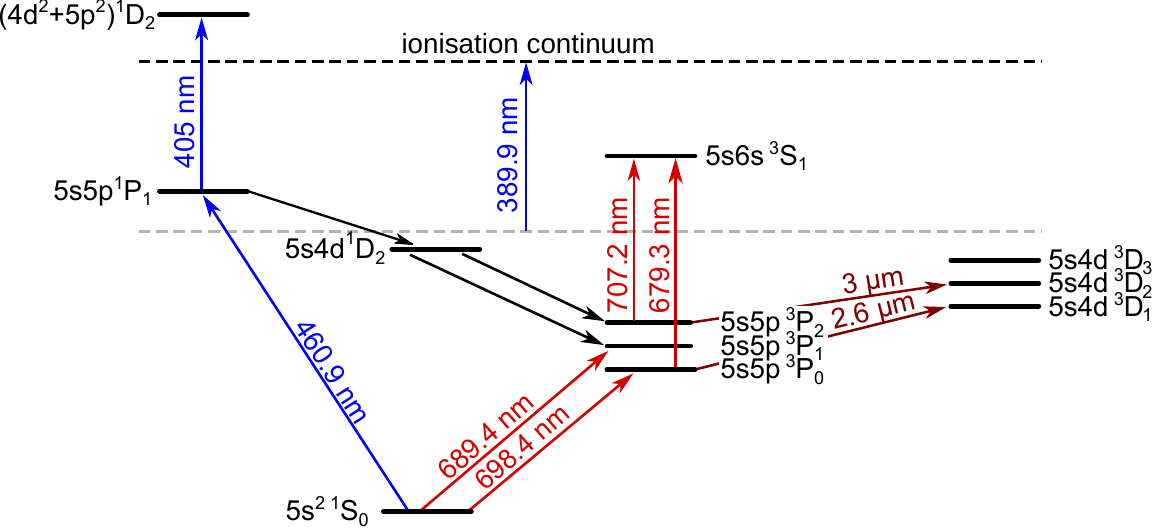}
 % \centering\includegraphics[width=0.6\columnwidth]{levels1.eps}
 \caption{Schematic energy level diagram showing relevant optical transitions used in the basic cycle of  a strontium optical lattice clock. The 460.9~nm transition is used in the first stage of cooling and for imaging of atoms. The  689.4~nm transition is used in the next cooling stages and the optical pumping in fermionic isotopes. The 707.2~nm and 679.3~nm transitions are used to repump the $^3\text{P}_0$ and $^3\text{P}_2$ states. The 698.4~nm transition is the clock transition. The 405~nm wavelength corresponds to the autoionisation resonance (4$d^2$+5$p^2$)$^1\text{D}_2$. The states belonging to the area bounded by the dashed lines are potentially affected by the photoionisation light at 389.9~nm. {The 3~\textmu m and 2.6~\textmu m transitions can be used as an alternative repumping scheme.}}
 %\textcolor{purple}{@MW: we should add the 3D1 state and its wavelengths to this diagram and add a 1-2 sentences about it, both here and in main text. DK}}
 
 \label{fig:levels}
 \end{figure}

A blue-detuned magic wavelength for the ${}^{1}$S${}_{0}$-${}^{3}$P${}_{0}$ clock transition in ${}^{87}$Sr was experimentally determined to be 389.889(9)~nm~\cite{Takamoto2009}. 
%\sout{While the clock transition in ${}^{87}$Sr is below photoionisation threshold, other transitions that are involved during normal operation of ${}^{87}$Sr optical lattice clock, for instance during an electro-shelving detection technique, are potentially affected by the photoionisation. In the majority of strontium optical lattice  clocks interrogation cycles [20-22]
%~\cite{Ludlow2015,Takamoto2005,Ludlow2006} 
%atoms pass through  $^1$P$_1$ and $^3$S$_1$ states that are susceptible to ionisation by the  blue 389.889~nm wavelength light~(see Fig.~\ref{fig:levels}).}
While these blue-detuned magic wavelength photons do not have sufficient energy to ionise the atoms directly from the strontium clock (${}^{1}$S${}_{0}$ and ${}^{3}$P${}_{0}$) states, other states that are involved during the normal operation of the strontium optical lattice clock are potentially affected by the photoionisation (see the region between the dashed lines in Fig.~\ref{fig:levels}).
For instance, 
{ the first stage of cooling down  the atoms before loading them into the optical lattice is generally based on the ${}^{1}$S${}_{0}$-${}^{1}$P${}_{1}$ transition (460.9 nm in Fig.~\ref{fig:levels}), and the repumping during   detection is based on the ${}^{3}$P${}_{0}$-${}^{3}$S${}_{1}$ and ${}^{3}$P${}_{2}$-${}^{3}$S${}_{1}$ transitions (679.3 nm  and 707.2 nm, respectively, in Fig.~\ref{fig:levels})}
~\cite{Ludlow2015,Takamoto2005,Ludlow2006}.

%\sout{To measure the excited fraction of the atoms into ${}^{3}$P${}_{0}$ after the clock interrogation~\cite{Ludlow2015,Takamoto2005,Ludlow2006}, atoms are required to be optically pumped into $^1$P$_1$ and $^3$S$_1$ states that are susceptible to ionisation by the  blue 389.889~nm wavelength light~(see Fig.~\ref{fig:levels})}

To determine photoionisation cross sections for {$^1\text{P}_1$ and $^3\text{S}_1$} levels, we used two different experimental methods. To investigate the photoionisation effect on the $^1$P$_1$ state, we compared the dynamics of loading atoms into %\sout{to} 
the magneto-optical trap (MOT) with and without the  ionising blue-detuned magic wavelength 389.889~nm light, alternately. 
To measure the photoionisation cross section for $^3$S$_1$ state, 
we detect atomic losses from the optical lattice trap induced by the presence of an ionisation laser beam at blue-detuned magic wavelength
 during the strontium optical lattice clock operation. The results of both measurements were used to analyse how the photoionisation by the blue-detuned magic wavelength light affects the performance of the optical atomic clocks under typical experimental conditions. Since the magic wavelength does not depend heavily on the particular isotope, as was shown for the red-detuned magic wavelength of Sr at 813~nm \cite{Takano2017}, we based our research on a more abundant bosonic $^{88}$Sr isotope without limiting the generality of our results.
 
\section{Experiment}

\subsection{Experimental setup}

    The photoionisation experiment has been performed on the Sr optical lattice atomic clock setup described in detail in~\cite{Bober2015} operating on ${}^{88}$Sr bosonic isotope. A basic cycle of this clock consists of cooling and loading atoms into a 1D red-detuned optical lattice trap, an interrogation of the clock transition by an ultra-stable clock laser, and detection of the resulting atomic population in the ground ($^1\text{S}_0$) and the excited ($^3\text{P}_0$) clock states~\cite{Ludlow2015}. Cooling and loading of atoms into the optical lattice trap are achieved by two consecutive MOTs, operating on the $^1\text{S}_0$-$^1\text{P}_1$  (blue MOT) and $^1\text{S}_0$-$^3\text{P}_1$  (red MOT) transitions, respectively. The clock $^1\text{S}_0$-$^3\text{P}_0$ transition is interrogated by the $\pi$-pulse Rabi excitation. The clock cycle is concluded by the detection phase that measures the ratio of populations in $^1\text{S}_0$ and $^3\text{P}_0$ states with the help of an electron shelving scheme~\cite{Dehmelt1982}, including fluorescence imaging on  $^1\text{S}_0$-$^1\text{P}_1$ transition and optical pumping on $^3\text{P}_{0}$-$^3\text{S}_1$ and $^3\text{P}_{2}$-$^3\text{S}_1$ transitions.

The photoionisation laser beam is synthesised via frequency doubling of a TiSa tunable laser light inside a bow-tie enhancement cavity. The resulting laser beam is spatially filtered with polarisation-maintaining single-mode fibre and expanded by sets of lenses to either $\sigma_x$=7.034(85)~mm and $\sigma_y$=6.07(30)~mm or $\sigma_x$=2.19(11)~mm and $\sigma_y$=1.88(12)~mm beam waist radii and directed 
%\textcolor{green}{
on
%} 
%\textcolor{green}{\sout{into}} 
atoms trapped by the blue MOT or the optical lattice trap, respectively. 
%\textcolor{green}{ with the larger beam used in $^1\text{P}_1$ photoionisation measurements and smaller one in $^3\text{S}_1$ measurements}. 
%In both cases, 
The diameter of the ionisation beam is much larger than the  typical dimensions of atomic clouds, i.e., 2.4~mm $\times$ 2.4~mm and 157~\textmu m $\times$ 66~\textmu m for blue MOT and the optical lattice, respectively. The frequency of the ionising laser is stabilised to a wavemeter through an analogue feedback loop with an accuracy %\sout{of} 
better than 100~MHz. 

\subsection{Photoionisation of the $^1\text{P}_1$} % (fold)

To determine photo-induced losses from the $^1$P$_1$ state due to blue-detuned magic 389.889~nm light, we analysed the fluorescence signal at 461~nm emitted by $^{88}$Sr atoms during  the blue MOT loading phase. To ensure that the $^1\text{P}_1$ is the only possible ionised state, the repumping laser beams for the $^3\text{P}_0$ and $^3\text{P}_2$ states were switched off (see Fig.~\ref{fig:levels}). The photoionisation cross section $\sigma_{^1\text{P}_1}$ was determined by comparing the loading rates of the MOT fluorescence with and without the photoionising 389.889~nm light. This technique has proven very efficient in measuring absolute ionisation cross sections of trapped atoms~\cite{Dinneen1992, Witkowski2018}.

The rate equation for the number of atoms  $N_\text{Sr}$ loaded into the MOT can be expressed as~\cite{GABBANINI199725} 

\begin{equation}
\frac{d N_\text{Sr}}{dt} = L_\text{Sr}- \left(\gamma_\text{Sr} + \gamma_P\right) N_\text{Sr} - \beta_\text{SrSr}\int dr^3 n^2_\text{Sr}, \label{eq:rate_full}
\end{equation}

\noindent
where $L_\text{Sr}$ is the  MOT loading rate, $\gamma_\text{Sr}$ is the combined loss  coefficient due to collisions with background gases, optical pumping to the metastable states and other possible single atom losses, $\gamma_P$ is the loss coefficient due to photoionisation, $\beta_\text{SrSr}$ is the loss coefficient due to light-assisted collisions between Sr atoms, and $n_\text{Sr}$ is the spatial density of  trapped atoms. In this approach, photoionisation is considered as another mechanism of  losses, linearly dependent on the number of atoms. The approach  is valid  if the photoionising beam does not modify the density distribution of the MOT (e.g. by a dipole force), which is always fulfilled in our setup.

The experiment was performed in the low-density regime,
{which means that the mean free path of the atoms is larger than the size of the trapped atomic cloud. A typical number of strontium atoms in a  blue MOT of a diameter of $\sim$2~mm is $\sim$6$\times$10$^8$~\cite{Bober2015}. It gives an atomic density of $\sim$10$^{11}$~cm$^{-3}$. Using the $^{88}$Sr collision cross-section of 10$^{-13}$~cm$^2$~\cite{Ferrari06}, the mean free path of the atoms trapped in the blue MOT is $\sim$100~cm, which is very large compared to the size of the MOT.}
%~\footnote{\textcolor{blue}{By low density regime we mean that the mean free path of the atoms is very large compared to the size of the trapped atomic cloud. In this experiment the number of strontium atoms trapped inside the spherical blue MOT of diameter 2.4 mm is $\approx$~6$\times$~10$^8$~\cite{Bober2015}. So, the calculated atomic density (n) inside the atomic cloud is $\approx$~8.3$\times$10$^{10}$~cm$^{-3}$. Using the measured $^{88}$Sr collision cross-section ($\sigma$) of 10$^{-13}$~cm$^2$~\cite{Ferrari06}, the calculated mean free path of the atoms trapped inside the blue MOT is $\approx$~80~cm, which is very large compared to the size of the atomic cloud.}}, 
The low-density regime enables us to neglect
%\textcolor{green}{us to ignore} %\textcolor{green}{\sout{skipping}}  
the last term in Eq.~(\ref{eq:rate_full}) as the  collisions  between trapped atoms are negligible. This yields the simplified rate equation

\begin{equation}
\frac{d N_\text{Sr}}{dt} = L_\text{Sr}- \left(\gamma_\text{Sr} + \gamma_P\right) N_\text{Sr}. \label{eq:rate_simplified}
\end{equation}

\noindent
Integration of Eq.~(\ref{eq:rate_simplified}) over time gives the formula for the  dependence of the number of atoms on time during MOT loading

\begin{equation}
N_\text{Sr}(t) = \frac{L_\text{Sr}}{\gamma_\text{Sr} + \gamma_P}\left(1-\exp\left(-\left(\gamma_\text{Sr} + \gamma_P\right) t \right)\right) \label{eq:sol_loadSr},
\end{equation}

\noindent
where  $N_\text{Sr}(t=0)=0$.

\begin{figure}[hbt]
    \centering\includegraphics[width=7cm]{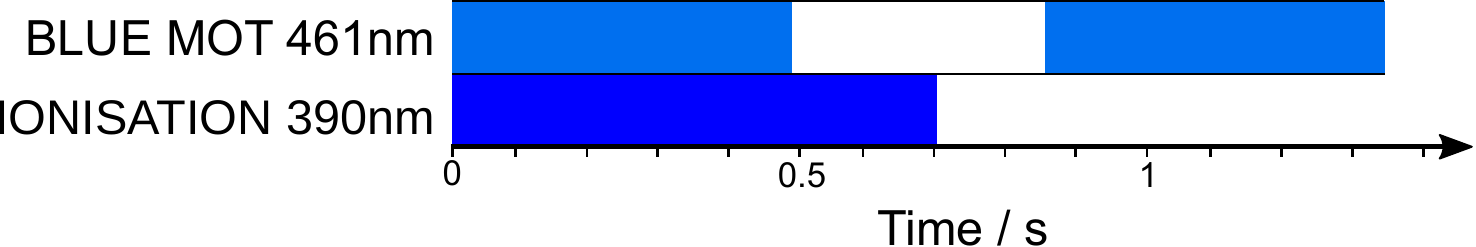}
\caption{The timing sequence of the lasers used in the $^1$P$_1$ photoionisation experiment. The time of the blue MOT's loading phase is identical in each sequence. To record the background from the ionisation laser beam, it was kept on for a 200~ms longer compared to MOT's loading phase duration.}
\label{fig:timeline_1P1}
\end{figure}

\begin{figure}[hbt]
\centering\includegraphics[width=7cm]{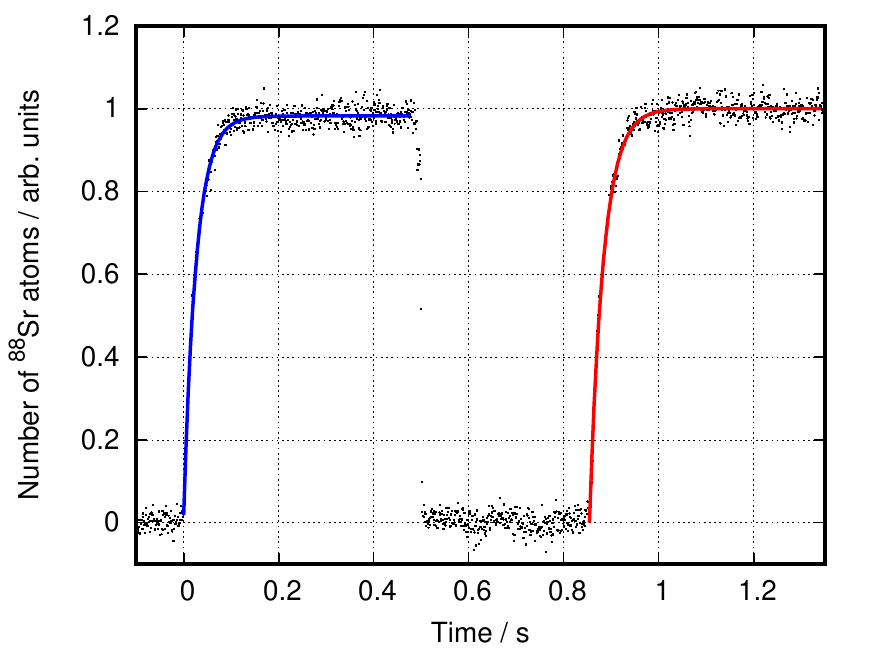}
% \centering\includegraphics[width=7cm]{loading_curves_of_Sr.eps}
\caption{Typical fluorescence of $^{88}$Sr atoms detected while loading into the blue MOT in the presence  (left) and in the absence (right) of the photoionising 389.889~nm light. The solid  blue and red  lines  depict fitted Eq.~(\ref{eq:sol_loadSr}).}
\label{fig:loading_curves}
\end{figure}

The loss coefficient $\gamma_P$ is related to the intensity $I_P$ of the ionising light through the expression

\begin{equation}
  \gamma_P  = \rho_{^1\text{P}_1}\sigma_{^1\text{P}_1} \frac{I_P}{h\nu_P}\label{eq:gamma_P},
\end{equation}

\noindent
where $\rho_{^1\text{P}_1}$ is the fraction of atoms in $^1\text{P}_1$ excited state, $\sigma_{^1\text{P}_1}$ is the photoionisation cross section, and $I_P/h\nu_P$ is the ionising photon flux. {The frequency $\nu_P$ is determined with the {type B standard } uncertainty $u(\nu_P)$ {derived from} the accuracy of the wavemeter}. To find the value of the $\gamma_P$ coefficient, two blue MOT loading curves were recorded sequentially in the presence and absence of the photoionisation laser beam (see Fig.~\ref{fig:timeline_1P1}). By fitting  Eq.~(\ref{eq:sol_loadSr}) independently to  both curves, we obtained loading rates $\gamma_\text{Sr}$ and $\gamma_\text{Sr}+\gamma_P,$ and determined $\gamma_P$ by their difference. The typical loading curves detected in the experiment  are presented in Fig.~\ref{fig:loading_curves}. For these curves, the loss coefficients are $\gamma_\text{Sr}$=34.08(80)~s$^{-1}$ and $\gamma_\text{Sr}+\gamma_P$=37.9(1.5)~s$^{-1}$, which corresponds to the reduction of the 1/e MOT loading time from 29.34(69)~ms to {26.4(1.0)~ms.} To block the scattered light from the ionisation laser, the blue MOT fluorescence was filtered by a 461~nm interference filter and then focused on a photodiode.
\added[id=mw,comment="I do not like the reviewer's suggestion to join Fig. 2 and 3 together"]{}

To determine the fraction of atoms in the excited state $\rho_{^1\text{P}_1},$ a model for two-level system was used

\begin{equation}
  \rho_{{}^{1}\text{P}_{1}} = \frac{1}{2}\frac{{I_{461}/I_{sat}}}
  {{I_{461}/I_{sat}}
  +4{\left(\Delta/\Gamma\right)^2}+1},\label{eq:fraction}
\end{equation}

\noindent
where $I_{461}$ is the total  intensity of the blue MOT trapping laser light, $I_{sat}$ is the  $^1\text{S}_0$-$^1\text{P}_1$ transition saturation intensity (427~W/m$^2$), $\Gamma$~is the natural decay rate of the excited ${}^1$P${}_1$ state ($2\pi\times 32$~MHz) and $\Delta$~is the trapping laser detuning from resonance (typically 2$\Gamma$). The population fraction of the atoms in ${}^1$P${}_1$ state was varied by changing the trapping laser intensity. To calibrate the trapping laser intensity, a small portion of the trapping 461~nm laser light was uncoupled before splitting and sent to a photodetector to monitor the intensity $I_{461}$.

The detuning $\Delta$ was continuously measured by an optical frequency comb.
{The related uncertainties $u(I_{461})$ and}
{$u(\Delta)$ {were} determined as  standard deviations of their means}. 
The intensity of the trapping laser light and its detuning from the resonance were continuously recorded by data acquisition software. 
\deleted[id=mw]{Fig.~%\ref{fig:PIvMOT}  
depicts the measured loss rate versus blue MOT trapping light power.} 
The trapping beam is expanded to a diameter of 2~cm and is split into three retro-reflected beams. The typical value of the total  intensity of the blue MOT trapping laser light seen by atoms, taking into account losses on viewports and optics, is  $I_{461} = 6\times $30~W/m${}^2 \approx 0.42~I_{sat}$.
\deleted[id=mw,comment="I removed Fig. 4"]{Since the linear fit is an approximation that is correct only for low powers,
and} \added[id=mw]{The} {methods that calibrate population dependence on the MOT beams intensity,} described e.g. in \cite{Shah07,Witkowski2018}, \added[id=mw]{are not used} {here since they} require laser light {intensities} exceeding the saturation of the transition. \deleted[id=mw]{the plot is not used for the population calibration and instead the} \added[id=mw]{Instead, the }measured values of $I_{461}$ and $\Delta$ are directly used in Eq.~(\ref{eq:fraction}).

%\begin{figure}[hbt]
%%\centering\includegraphics[width=7cm]{nice1.pdf}
%\centering\includegraphics[width=7cm]{ugly3.eps}
%\caption{The measured loss rate versus blue MOT trapping light power at $\Delta = 2\Gamma$. The linear fit is an approximation that is correct only for low powers, therefore it was not used in the calculations.}
%\label{fig:PIvMOT}
%\end{figure}

\subsection{Photoionisation of the $^3\text{S}_1$} % (fold)

\begin{figure}[hbt]
\centering\includegraphics[width=0.8\columnwidth]{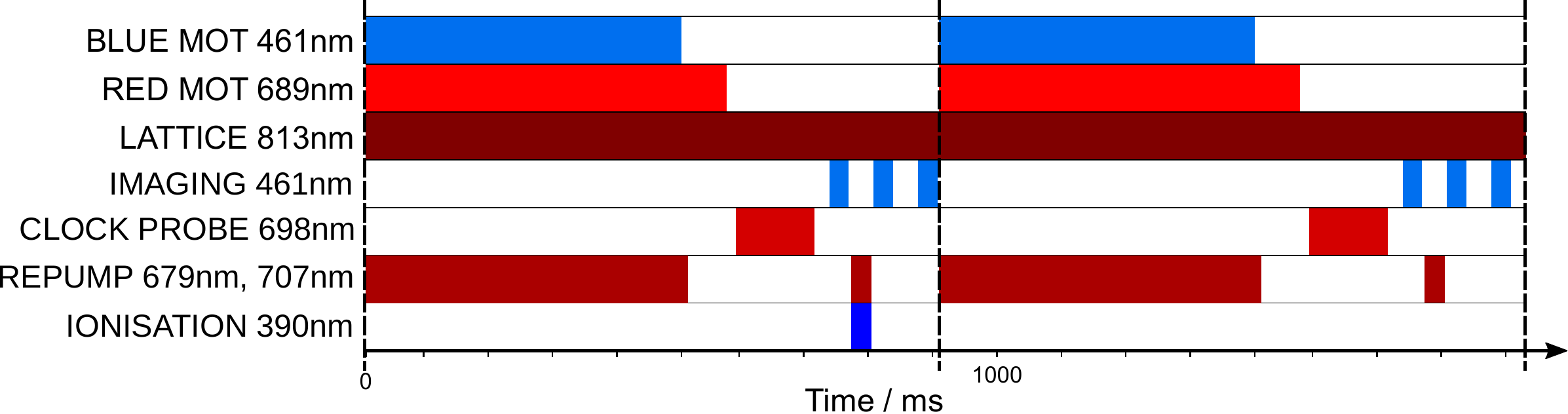}
\caption{The timing sequence of the lasers used in the $^3$S$_1$ photoionisation experiment. 
It consists of two clock cycles, the first one (left) with an added photoionising pulse, the second one (right) providing a background.
\label{fig:timeline_3S1}}
\end{figure}

\added[id=mw, comment="Reviewer mentioned that a sentence was lost at the beginning of 2.3. This was the sentence about the transition probability that we decided to be discarded."]{}To measure the photoionisation cross section of $^3\text{S}_1$ state at 389.889~nm, we employed a different experimental approach. The experiment was performed during the standard clock cycle of the strontium optical lattice clock described in detail elsewhere~\cite{Bober2015}. %\textcolor{blue}{The photoionisation cross section was directly inferred from the photoionisation-induced reduction in the excited-state $^3\text{P}_0$ population %number of atoms in the excited state $\Delta N_e=N_e^I-N_e$, }
%change in transition} probability $P$ of the $^1\text{S}_0$-$^3\text{P}_0$ \textcolor{blue}{clock} transition  
%\begin{equation}
%  P=\frac{N_e}{N_g + N_e},
%  \label{eq:prob}
%\end{equation}
%\noindent
%\textcolor{blue}{where $N_e^I$ and $N_e$ are populations in the excited $^3\text{S}_1$ state with and without the presence of the photoionisation beam, respectively.} 
The timing sequence of the laser beams used in the experiment is shown in Fig.~\ref{fig:timeline_3S1}. After %\textcolor{green}{
sequential
%} %\textcolor{green}{\sout{subsequent}} 
cooling in  blue and red MOT  Sr, atoms are transferred into the 1D vertical red-detuned magic wavelength optical lattice at 813 nm and interrogated by an ultrastable clock laser. The clock laser is locked to the $^1\text{S}_0$-$^3\text{P}_0$ transition in the second, independent strontium optical lattice clock. 
%\sout{The clock laser pulse is long in comparison to the transition Rabi frequency. 
%As a result, one-half of atoms ($N_e$)  can be  excited to the $^3\text{P}_0$ state while the rest of atoms ($N_g$) remains in the ground state.} 
The clock laser interrogation pulse, long with respect to the transition Rabi frequency, is applied exactly on the transition centre frequency and excites one-half of the atoms ($N_e$)   to the $^3\text{P}_0$ state while the rest of the atoms ($N_g$) remain in the ground  $^1\text{S}_0$ state.
Subsequently,
%\sout{After the \textcolor{blue}{interrogation by the clock laser},} the imaging pulse at 461~nm is applied to the ground-state atoms and the population $N_g$ is  destructively \textcolor{red}{counted}\sout{measured} by fluorescence detection with the electron-multiplying CCD. 
after that, %\sout{Subsequently,} 
Sr atoms initially excited to the $^3\text{P}_0$ state are pumped back to the ground state through the $^3\text{S}_1$, $^3\text{P}_1$, and $^3\text{P}_2$ states with 679.3~nm and 707.2~nm transitions (see Figs.~\ref{fig:levels} and~\ref{fig:timeline_3S1}). The number of previously excited atoms $N_e$ is then determined by the second imaging pulse at 461~nm. The detection is concluded by a %\sout{ followed by the} 
background image of the empty trap.

A pulse of the photoionising beam added during the repumping phase of the first clock cycle opens a new channel of losses due to photoionisation from the ${}^3$S${}_1$ state. Therefore, it decreases repumping efficiency and, consequently, reduces the number of atoms recorded by the second imaging pulse. 
The photoionisation cross section $\sigma_{^3\text{S}_1}$ is  inferred from the photoionisation-induced reduction in the number of atoms %excited-state $^3\text{P}_0$ population 
$\Delta N_e=N_e-N_e^I$, where $N_e^I$ and $N_e$ are the excited-state $^3\text{P}_0$ populations detected sequentially in the presence (the first cycle in Fig.~\ref{fig:timeline_3S1}) and absence of the photoionisation beam (the second cycle in Fig.~\ref{fig:timeline_3S1}), respectively. 

%The absolute value of the photoionisation cross section $\sigma_{^3\text{S}_1}$ was deduced from observed transition probabilities detected sequentially }
%\textcolor{green}{
%in
%} 
%\sout{with} 
%the presence (the first cycle in Fig.~\ref{fig:timeline_3S1}) and 
%\textcolor{green}{\sout{the}} 
%absence of the photoionisation beam (the second cycle in Fig.~\ref{fig:timeline_3S1}). 

\section{Results}
\subsection{Photoionisation cross section at blue magic wavelength} % (fold)

The photoionisation cross section $\sigma_{^1\text{P}_1}$  was determined from the loss coefficient $\gamma_P$ and characteristics of both the blue MOT and the ionising beam. The atomic distribution in the blue MOT detected by a CCD camera is well described by the Gaussian-like profile

\begin{equation}
  N(x,y)=\frac{2N_0}{\pi r_x r_y}\exp\left( \frac{-2x^2}{r_x^2} \right)\exp\left( \frac{-2y^2}{r_y^2}\right),\label{eq:N_distribution}
\end{equation}

\noindent
where $r_x,r_y$ are MOT radii (about 1.2~mm, typically). The averaged intensity of the ionising light $\langle I_P \rangle$ seen by atoms in the blue MOT is given by

\begin{equation}
  \langle I_P \rangle =\int \int \frac{I(x,y)N(x,y)}{N_0} dx dy,\label{eq:avr_intensity}
\end{equation}

\noindent
where $N_0$ is the total number of atoms in the MOT, and $I(x,y)$ and $N(x,y)$ are distributions of the intensity of the beam and the atoms, respectively. The intensity profile of the photoionising beam was determined by a CCD beam profiler measurement and corrected with the  coefficient of transmission through the vacuum chamber viewports. {The intensity of the ionising light was power-locked and continuously monitored by a photodiode, and the related uncertainty $u(I_P)$ was calculated as a standard error of the mean of the recorded photodiode readout. The photodiode was calibrated based on the uncertainties related to the atomic distribution, namely $u(r_x)$ and $u(r_y)$. } 

The photoionisation cross section $\sigma_{^1\text{P}_1}$ was calculated according to Eq.~(\ref{eq:gamma_P})  by combining the measured values of the excited state fraction $\rho_{^1\text{P}_1}$, loss coefficient $\gamma_P$, and the intensity $\langle I_P \rangle$, monitored and recorded during measurements. {We recorded 12 to 20 pairs of MOT loading curves sequentially with (the first curve) and without the photoionisation beam (the second curve) for each wavelength of the photoionisation beam separately. For each pair of curves, we determined the ionisation loss rate $\gamma_P$ as the difference $(\gamma_{Sr}+\gamma_P)-\gamma_{Sr}$, where $\gamma_{Sr}+\gamma_P$ and $\gamma_{Sr}$ were determined by fitting the Eq.~(\ref{eq:sol_loadSr}) to the first or the second curve, respectively. The uncertainty $u(\gamma_P)$ was calculated independently for each wavelength as the standard deviation of the mean. {Finally, t}he uncertainty of the photoionisation cross section $u(\sigma_{^1P_1})$ was calculated as a combined standard uncertainty with several independent quantities, namely $u(\gamma_P)$, $u(I_P)$, $u(\Delta)$, $u(I_{461})$, and $u(\nu_p)$.}
%Marcin, widzisz czat?
%Oprocz tego, to samo trzeba zrobic dla 1S3

\begin{figure}[hbt]
\centering\includegraphics[width=6cm]{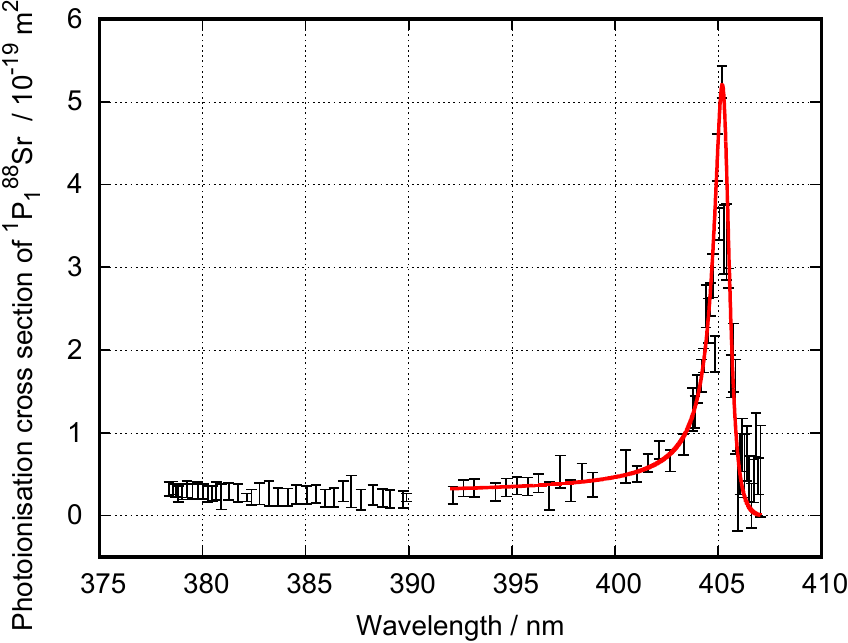}
\caption{The photoionisation cross section  from the $^1\text{P}_1$~state as a function of the wavelength of the ionising light. The solid red line is a Fano profile fitted to the measured data. The fitting range is limited to the experimental points above 392~nm to exclude  another possible resonance below 389.9~nm.}
\label{fig:results}
\end{figure}

To verify our experimental method, we extended the range of photoionisation wavelengths to cover the autoionisation resonance $(4d^2+5p^2)^1\text{D}_2$ near 405~nm \cite{Ewart1976,Esherick1977,Baig1998}. This way, we enabled comparison with previous measurements of the autoionisation resonance performed with different experimental approaches~\cite{Mende1995, Sami-ul-Haq2006}.
The measurements of $\sigma_{^1\text{P}_1}$ were carried out for wavelengths in a range from 378.4~nm to 407~nm. The results are shown in Fig.~\ref{fig:results}. %Each point is an averaged value of %\textcolor{orange}{\textbf{several dozen}} 
%at least twelve measurements with an uncertainty estimated as a mean standard deviation.  

\noindent

By fitting a Fano profile~\cite{Fano1961} to the measured points, we determined the position of the resonance and its peak value to be $\lambda_R$=405.196(44)~nm and $\sigma_R$=5.20(94)$\times$10$^{-19}$~m$^2$, respectively. To exclude  another possible resonance observed on the high energy wing of the measured photoionisation spectra, the fitting range was limited to the experimental points above 392~nm. The gap in results between 390~nm and 392~nm is due to technical problems related to the locking of TiSa laser within the corresponding range. Our results are consistent with the results presented previously by Mende et al. ($\lambda_R$=405.213(2)~nm, $\sigma_R$ = 5.6(1.1)$\times$10$^{-19}$~m$^2$)~\cite{Mende1995} and by Sami-ul-Haq et al. ($\lambda_R$= 405.177(16)~nm, $\sigma_R$=5.45(98)$\times$10$^{-19}$~m$^2$)~\cite{Sami-ul-Haq2006}. Note that previous measurements, unlike ours, did not involve cold atomic samples.  

At the blue magic 389.889~nm 
%\textcolor{green}{
wavelength,
%} 
the photoionisation cross section is equal to 2.20(50)$\times$10$^{-20}$~m$^2$ (see Fig.~\ref{fig:results}).  
Our result is consistent with the value deduced from the curve reported by Mende et al.~\cite{Mende1995} 1.46(29)$\times$10$^{-20}$~m$^2$, within the 20\% uncertainty claimed by the author. In our case, the most significant factor contributing to the uncertainty of $\sigma_{^1\text{P}_1}$ comes from the fluctuation of the number of atoms loaded into the blue MOT. 
%This fluctuation directly translates into the fitting error of the loss coefficient and then on the uncertainty of the photoionisation cross section.  
{The atom number fluctuation affects the fitting uncertainty, however, as the statistical error, it is averaged, and it does not contribute to the final result, only to its uncertainty.}  

In the second set of experiments, the photoionisation cross section from the $5s6s  ^3\text{S}_1$ excited state was determined by %\color[blue]{}
%comparing the \textcolor{blue}{transition probabilities of the atoms from} $^1\text{S}_0$ to $^3$P${}_0$ \textcolor{blue}{in the presence} ($P^I$) and \textcolor{blue}{absence} ($P$) of the photoionising 389.889~nm light during the repumping phase. Since the photoionisation-induced  atomic losses are very small ($\Delta N_e/N_e \ll 1$), the ratio $P/P^I$ can be written as a series expansion 

\begin{equation}
    \sigma_{^3\text{S}_1}=\frac{\left <\Delta N_e\right>}{\left<N_e\right>}  \frac{h\nu_P}{\langle I_P \rangle t_\text{int}} \label{eq:sigma3s1},
\end{equation}

\noindent
%\textcolor{orange}{\it the meaning of average <> for N is not defined}
where $\left <N_e\right >$ and $\left <\Delta N_e\right >$ are the averaged excited state $^3\text{P}_0$ population and its averaged photoionisation-induced reduction during the repumping through the $^3\text{S}_1$ state, respectively, $\langle I_P \rangle$ is the averaged intensity of the photoionisation radiation seen by atoms in the optical lattice calculated according to Eq.~(\ref{eq:avr_intensity}), and $t_\text{int}$ is the 
%effective 
{interaction} time {between photoionisation beam and}
%of the exposition to the ionising beam for 
the atoms in the $^3\text{S}_1$ state. The value of $t_\text{int}=$~43~ns was determined on the basis of 
%\textcolor{blue}{our repumping scheme, } 
both $^3\text{S}_1\rightarrow$ $^{3}\text{P}_2$ and $^3\text{S}_1\rightarrow$ $^{3}\text{P}_0$ transition probabilities \cite{Garcia1988} and the natural lifetime of the $^3\text{S}_1$ state \cite{Jonsson1984,Porsev2008}. {Additionally, in calculating the interaction time $t_\text{int}$, we included
%took into account the process of 
repumping to the excited $^3\text{S}_1$ state back from $^3\text{P}_2$ and $^3\text{P}_0$ states}. The ionisation beam was power-locked  with an intensity of 26270(80)~W/m$^2$ at the Gaussian-profile peak.
The $e^{-2}$ diameters of the beam (2.19(11)$\times$1.88(12)~mm) are %\sout{is}
much larger than the dimensions of the illuminated atomic cloud 
% \sout{of} \textcolor{red}{
(157(10)~\textmu m x 66(6)~\textmu m). 
 %   \begin{equation} 
  %  \frac{P}{P^I}=\frac{N_e}{N_g+N_e}\frac{N_g^I+N_e^I}{N_e^I}=\frac{N_e(N_g+N_e-\Delta N_e)}{(N_e-\Delta N_e)(N_g+N_e)}\approx 1 + \frac{N_g}{N_g+N_e}\frac{\Delta N_e}{N_e},
%\end{equation}
%\noindent
%where the  superscript $I$ denotes the quantities related to the photoionisation  and  $\Delta N=N_e-N_e^I$ is the difference between the excited state populations. %\sout{The experimental procedure ensures $N_g=N_g^I$.} 
To ensure %the  equality $N_g=N_g^I$ 
the stability of the number of atoms %is kept valid 
throughout each experimental cycle, we monitored both ground-state populations $N_g$ and $N_g^I$ independently. The experimental points satisfying the relation $(N_g-N_g^I)/N_g>2\%$ were excluded resulting in 3292 points left for further analysis. To be less prone to the oscillations of the number of atoms, we made the interleaved measurements randomly staggered by the manual triggering of  consecutive cycles. 
%The \textcolor{blue}{upper limit on the }atomic  losses  $\langle \Delta N_e/N_e \rangle$ averaged over 
%\textcolor{orange}{\textbf{several hundred}} 
%\textcolor{blue}{3292} measurements  was  measured to be 
%\textcolor{orange}{\textbf{less than}} 
%0.26\%. The photoionisation cross section $\sigma_{^3\text{S}_1}$ can thus be estimated as 

%\begin{equation}
%    \sigma_{^3\text{S}_1}=\left< \frac{\Delta N_e}{N_e} \right> %\frac{h\nu_P}{\langle I_P \rangle t} \label{eq:sigma3s1},
%\end{equation}

%\noindent
%where $\langle I_P \rangle$ is the averaged intensity of the photoionisation radiation seen by atoms in the optical lattice, calculated according to Eq.~(\ref{eq:avr_intensity}), and $t$ is the ionisation exposure time. The ionisation beam was power-locked  with an intensity of 26270(80)~W/m$^2$ at the Gaussian-profile peak.
%The $e^{-2}$ diameter\textcolor{blue}{s} of the beam \textcolor{blue}{ (2.19(11)$\times$1.88(12)~mm) are} %\sout{is}
%much larger than the dimensions of the illuminated atomic cloud 
% \sout{of} \textcolor{red}{
%(157(10)~$\mu$m x 66(6)~$\mu$m). 

The determination of the cross section $\sigma_{^3\text{S}_1}$ based on Eq.~(\ref{eq:sigma3s1}) yields %$\sigma_{^3\text{S}_1}=2.4(1.4)\times10^{-18}~\text{m}^2$. 
$\sigma_{^3\text{S}_1}=1.38(66)\times$10$^{-18}$~\text{m}$^2.$
%$\sigma_{^3\text{S}_1}=4.5(4.5)\times10^{-23}~\text{m}^2$. 
{The uncertainty of the $\sigma_{^3\text{S}_1}$
%photoionisation cross section of the 3S1 state 
was calculated as the combined standard uncertainty involving the uncertainties $u(\left<Ne\right>)$, $u(\left<\Delta Ne\right>)$, $u(\left<I_P\right>)$, and $u(\nu_P)$. The frequency type B uncertainty  $u(\nu_P)$ is derived from the  resolution of the wavemeter. All the other terms are calculated as the standard deviations of their means.}

\subsection{Study of feasibility: strontium optical clock with blue-detuned magic wavelength optical lattice}

In this section, we use our measured photoionisation cross sections of $^1\text{P}_1$ and $^3\text{S}_1$ states to estimate the impact of the loss of the atoms from these states due to the ionisation in a blue-detuned magic wavelength optical lattice  on an optical strontium clock operation and suggest possible mitigation measures. 

As mentioned above, in a blue-detuned magic wavelength optical lattice, atoms are confined at the minima of light intensity as opposed to a red-detuned magic wavelength optical lattice. However, the atoms cannot be trapped in a simple 1D blue-detuned magic wavelength lattice trap because they will escape along the radial directions. Confinement in all three orthogonal directions can be achieved, for instance, by a 3D optical lattice trap, {made up of three independent 1D optical lattices~\cite{Will2013}. Moreover, a 3D optical lattice will reduce the influence of interactions between atoms on the optical clock's accuracy~\cite{Campbell2017}.} 

To reach the accuracy goal of state-of-the-art optical lattice clocks, all frequency shifts connected with motional effects must be suppressed. When each atom is confined in a single lattice site to a region much smaller than the wavelength of the clock probing laser, and any tunnelling between sites is negligible, the Doppler and recoil shifts are suppressed, and the atoms are in the Lamb-Dicke regime~\cite{Dicke1953,Wineland1979}. This implies different limits on minimal  required potential depth depending on the direction of the probing of the clock transition by a clock laser beam, either horizontal or vertical. On the other hand, higher potential means higher losses due to photoionisation.

\subsubsection{Horizontal direction}
%units of the recoil energy   associated with the absorption or emission of a photon of the lattice light,  $E^{\text{rec}}= h*14$~kHz (DK check value and h or hbar!!!!) associated to the 390~nm lattice, reaching an uncertainty of  $10^{-18}$ requires a trap depth of at least 125~$E^{\text{rec}}$, similar to the 120~$E^{\text{rec}}$ required  for the 813~nm magic wavelength}

In a horizontal periodic potential, the
states with the same vibrational quantum number in different potential wells are degenerated in energy, amplifying tunnelling between the wells. This will spread out the spatial wave functions of the atoms so that they are not localised to a single well  and create a band structure in their energy spectrum. This yields,  depending on the initial state of the atoms in the trap, a broadening and a shift of the atomic transition of the order of the bandwidth of the lowest energy band of the system~\cite{Lemonde2005}. Figure~3 of~\cite{Lemonde2005}  shows the corresponding bandwidth with the bandwidth in
units of the recoil energy  $E^{\text{rec}}$ associated with the absorption or emission of a photon of the lattice light and in 
Hz units calculated for 813 nm magic wavelength.
%Figure \ref{fig:tunneling} depicts the dependence of the ground state bandwidth  on the lattice depth in Hz units calculated for blue-detuned  magic wavelength. 
%\textcolor{green}{\sout{Considering the potential depth $U_0$ in 
%units of the recoil energy   associated with the absorption or emission of a photon of the lattice light,  $E^{\text{rec}}= h*14$~kHz (DK check value and h or hbar!!!! \textcolor{red}{for 389.9 nm I get $E^{rec}=h*14.91$ ~kHz, so it h, not $\hbar$, but its 15 kHz.}) associated to the 390~nm lattice, 
%to avoid any line broadening of the atomic transition at the  $10^{-18}$ level
%\sout{reaching an uncertainty 
%of  $10^{-18}$}
%with horizontal direction of the clock transition probing requires a trap depth of at least 125~$E^{\text{rec}}$., similar to the 120~$E^{\text{rec}}$ required  for the 813~nm magic wavelength}}

%\sout{For instance, to reach the goal of a fractional uncertainty of the optical lattice clock below $10^{-18}$, the bandwidth should be below $\sim 1$~mHz. Figure~\ref{fig:tunneling} shows that for the ground state this requires the potential depth above $\sim 125~E^{\text{rec}}_i$. }
Since the recoil energy scales like $k^2$, where $k$ is the  wavenumber of the lattice light, in the blue-detuned lattice, for instance, the bandwidth of $\sim 1$~mHz, corresponding to the $10^{-17}-10^{-18}$ accuracy range, requires the potential depth above $\sim 125~E^{\text{rec}}.$

\subsubsection{Vertical direction}

\begin{figure}[hbt]
	 \centering
	 \includegraphics[width=0.95\columnwidth]{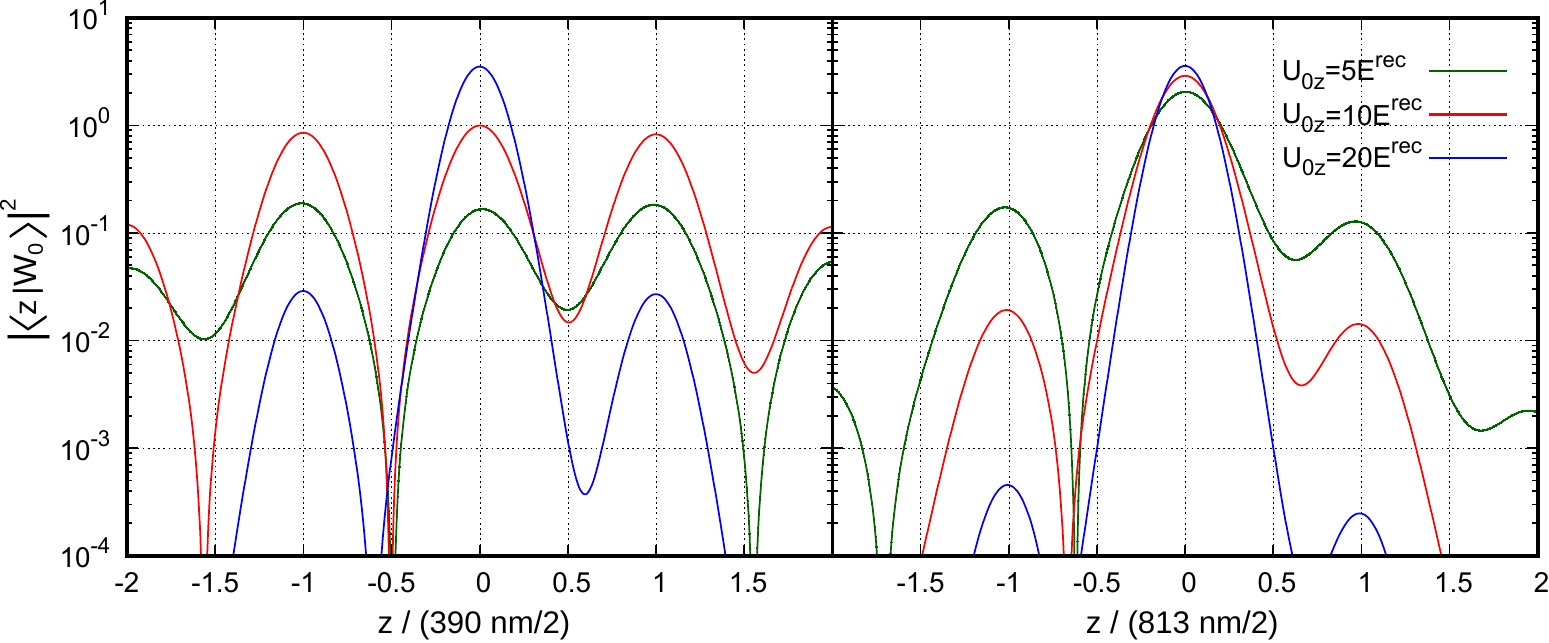}
	\caption{Wannier-Stark states in position representation for different lattice depths for 390~nm (left) and 813~nm (right) magic wavelength lattices.\label{fig:WS}}
\end{figure}

For a vertical clock laser probe beam, the approach from the previous subsection is no longer valid due to the presence of gravity. In the periodic potential in an accelerated frame, energies of atoms in adjacent lattice sites are shifted, and the Hamiltonian no longer supports bound states. This means that an atom in a vertical optical lattice will eventually tunnel to a continuum. Fortunately,  the timescale of this process increases exponentially with lattice depth~\cite{Lemonde2005}, and this is not an issue with the lattice depths considered here. Therefore, we must adopt %\sout{another approach for the treatment of atoms trapped in a vertical lattice -} 
the formalism of Wannier-Stark states~\cite{Wannier1960}. The external Hamiltonian of an atom with mass $m$ in a vertical lattice optical trap of the depth $U_{0z}$ is given as

\begin{equation}
  \hat{H}=\frac{\hbar^{2}\hat{\kappa}^{2}}{2m}+\frac{U_{0z}}{2}[1-\cos(2k\hat{z})]+mg\hat{z},
\label{eq:HWS}
\end{equation}

\noindent
where $\kappa$ is the atom quasimomentum in z-direction. %\textcolor{red}{$\kappa$ is the atoms' quasimomentum and is an operator/variable of the hamiltonian, and $k_{z}$ is the wavevector of the lattice and constant. }
%\sout{We assume $R_z=1$ and we use a simplified form of 1D potential of vertical optical lattice as given in [39]} %\cite{Lemonde2005}. 
The eigenstates of this Hamiltonian are called Wannier-Stark states $|W_M\rangle$, where $M$ denotes $m$th well of the lattice. 
We have constructed the Wannier-Stark state  as the sum of Wannier states in different lattice sites (i.e. the eigenstates of Hamiltonian in Eq.~(\ref{eq:HWS}) without gravity) where every Wannier function is weighed by the appropriate Bessel function ~\cite{Gluck2001}.
Fig.~\ref{fig:WS} shows the spatial representation of Wannier-Stark states  $|W_0\rangle$ centred around site $M=0$ for different values of  the lattice depth $U_{0z}$. The states are calculated for 390~nm blue-detuned lattice and, for comparison, for widely-used 813~nm red-detuned lattice.
In both cases, we assumed that the tunnelling between different energy bands is negligible, i.e. the atom can only tunnel between lattice sites of the ground energy band.   

As seen in Fig.~\ref{fig:WS}, for an 813~nm red-detuned optical lattice, the Wannier-Stark state consists of a main central peak and two smaller "revival" peaks in the adjacent wells, even for shallow lattices. These revival peaks decay quickly with increasing lattice depth, with revival peaks being hundred times smaller at lattice depths of $10~E^{\text{rec}}_z$ {and the wave function being practically localised to one lattice site}\cite{Lemonde2005}. To achieve similar ratios of the main and side peaks for a 390~nm blue-detuned lattice, the lattice should be at least twice as deep, i.e. $U_{0z}=20~E^{\text{rec}}$ because of the shorter distance between lattice sites and thus smaller energy shifts in the adjacent sites.

\begin{figure}[hbt]
	 \centering
	 \includegraphics[width=0.95\columnwidth]{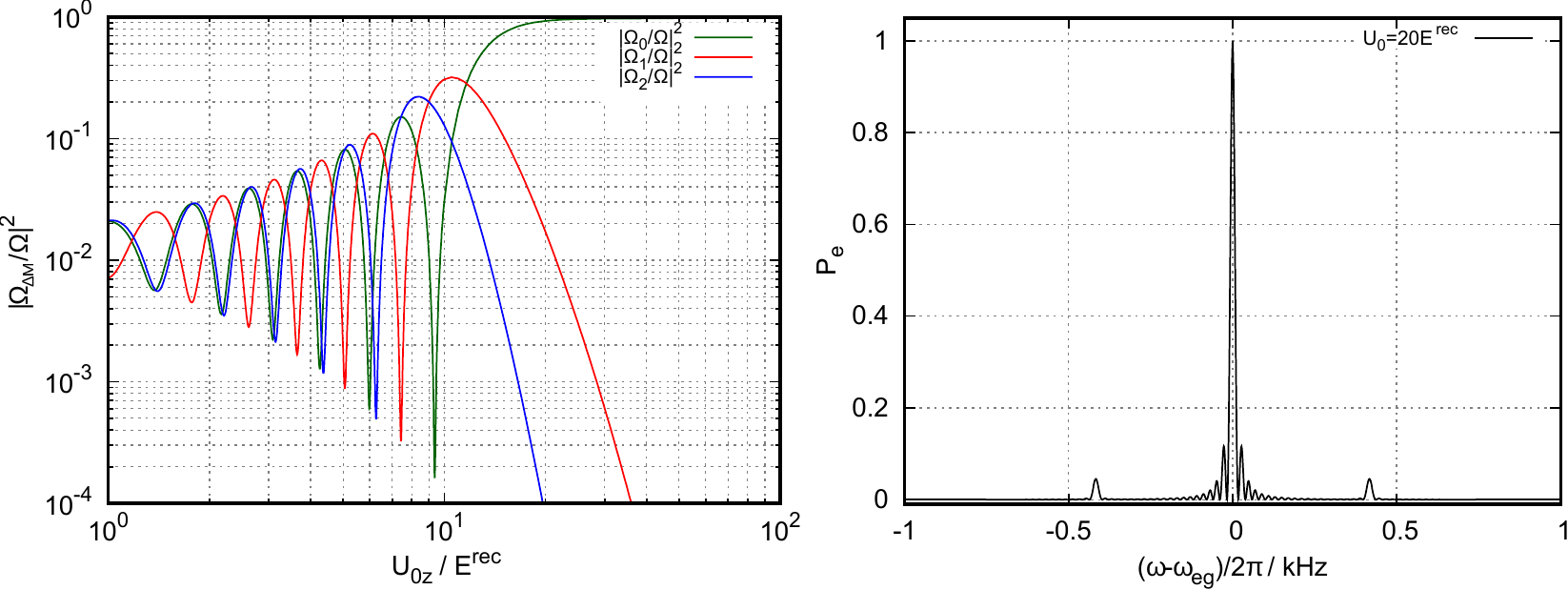}
		\caption{(left) Relative coupling strengths $|\frac{\Omega_{\Delta M}}{\Omega}|^{2}$ of the carrier ($\Delta M=0$)  and first 4 sidebands ($\Delta M=\pm 1$, $\Delta M=\pm 2$) as a function of lattice depth for a blue-detuned optical lattice. (right) Transition probability $P_e$ of the ${}^{1}$S${}_{0}$-${}^{3}$P${}_{0}$ clock transition for effective Rabi frequency $\Omega_{0}/2\pi=10$~Hz and $\pi$-pulse interaction time of 50~ms when the initial state is a pure Wannier-Stark state and $U_{0}=20~E^{\text{rec}}$. 
	%The sideband resonances are shifted by $\Delta_{g}/2\pi$ from the carrier frequency (right).  
		\label{fig:WS_coupling_390nm}}
\end{figure}

To examine the effects of coupling the Wannier-Stark states to their neighbours by the $^1$S$_0$-$^3$P$_0$ probe beam,
%on the Wannier-Stark states, 
we consider the Wannier-Stark ladder of states with an internal two-level energy structure  $\left( |g,W_M\rangle, |e,W_M^{'}\rangle\right)$. The Wannier-Stark ladder is a set of Wannier-Stark states with one $|W_M\rangle$ state in each lattice site. Wannier-Stark states in the adjacent lattice sites are separated in energy by $\hbar\Delta_g$ corresponding to the change in gravitational potential between sites.  For Sr, this separation between adjacent sites is equal to $\frac{\Delta_{g}}{2\pi}=417$~Hz for 390~nm blue-detuned lattice.
The energy difference between the ground $|g\rangle$ state and the excited $|e\rangle$ state of the atom is given by $\hbar\omega_{eg}$. The probe beam couples the $|g,W_M\rangle$ and $|e,W_M^{'}\rangle$ states with different coupling strengths $\Omega_{\Delta M}$,
where  $\Delta M=M^{'}-M=0$ and $\Delta M\neq0$ for coupling of WS states in the same lattice site and for coupling to neighbouring lattice sites, respectively.
%with $\Delta M=M^{'}-M=0$ for coupling of WS states in the same lattice site and $\Delta M\neq0$ for coupling to neighbouring sites. 
This coupling corresponds to a translation in momentum space $e^{ik_{c}z}$ \cite{Lemonde2005}:

\begin{equation}
    \Omega_{\Delta M}=\Omega\langle W_{M}|e^{ik_{c}\hat{z}}|W_{M^{'}}\rangle,
    \label{eq:WS_coupling}
\end{equation}

\noindent
where $k_{c}$ is the wavevector of the coupling probe laser and $\Omega$ is the Rabi frequency.
The relative coupling strengths $|\frac{\Omega_{\Delta M}}{\Omega}|^{2}$ of the "carrier" $\Omega_{0}$ and the first 4 "sidebands" $\Omega_{\pm 1,\pm 2}$ as a function of lattice depth are shown in Fig.~\ref{fig:WS_coupling_390nm} (left). For very shallow optical lattices with depths below
%$U_{0}=15 E_{z}^{\text{rec}}$ 
$15 E^{\text{rec}}$ 
the couplings show strong oscillations similar to those in \cite{Tackmann2011}. For higher lattice depths, couplings to neighbouring lattice sites rapidly decay and the atom becomes trapped in a single lattice with strong suppression of tunnelling between sites. {Additionally, Fig.~\ref{fig:WS_coupling_390nm} (left) shows that for shallow lattice it is possible to choose different ratios of coupling parameters $\frac{\Omega_{0}}{\Omega_{1,2}}$ by tight control of lattice depth. This control over coupling parameters allows engineering of the extent of atomic wavefunctions through the adjustment of trap depth and thereby lowering the collisional frequency shifts arising the on-site p-wave and neighbouring-site s-wave interactions, as done for ~813 nm optical lattice\cite{Bothwell2022}.}
For lattice depth of $20 E^{\text{rec}}$, coupling strengths to the nearest and second-nearest lattice site are $10^{-2}$ and $10^{-4}$ times weaker than coupling between the ground $|g,W_M\rangle$ and excited $|g,W_M\rangle$ WS state in the same lattice site.   
%\begin{figure}[hbt]
%	 \centering
%		\includegraphics[width=0.75\columnwidth]{WS_ladder_of_states.JPG}
%	\caption{\textcolor{green}{Wannier-Stark ladder of states and couplings between the ground and excited states in different lattice sites where $\omega_{eg}$ is frequency of the clock transition and $\hbar\Delta_{g}$ is the change in gravitational potential between adjacent lattice sites. 
%	\label{fig:WS_ladder}}}
%\end{figure}

To calculate the populations of the ground and excited WS states, we considered the evolution of the different states under coupling to the probe laser by numerically solving the set  of differential equations \cite{Lemonde2005}

\begin{equation}
\begin{aligned}
    i\dot a^{g}_{M}=\sum_{M^{'}}\frac{\Omega_{M-M^{'}}^{*}}{2}e^{-i\pi M^{'}k_{c}/k}e^{i\Delta_{M-M^{'}}t} a^{e}_{{M}^{'}}, \\
    i\dot a^{e}_{M}=\sum_{M^{'}}\frac{\Omega_{M^{'}-M}}{2}e^{i\pi M k_{c}/k}e^{-i\Delta_{M^{'}-M}t} a^{g}_{{M}^{'}}, 
    \label{eq:pop_evolution}
\end{aligned}
\end{equation}

\noindent
where $a^{g}_{M}$ and $a^{e}_{M}$ are the probability amplitudes of the ground and excited state, respectively, and $\Delta_{M-M^{'}}=\omega-\omega_{eg}+(M-M^{'})\Delta_g$. In these calculations we assume that the atom can only tunnel to the nearest lattice site and the initial state is a pure WS state.
%The transition probability of the excited WS state in a blue-detuned optical lattice as a function of detuning from resonance of the clock transition is given on Fig~\ref{fig:WS_coupling_390nm} (right). 

The computed resonances for 
%The numerical calculations were performed for 
a blue-detuned vertical optical lattice with depth of $20 E^{\text{rec}}$, an effective Rabi frequency $\Omega_{0}/2\pi=10$~Hz and interaction time of 50~ms
are shown in Fig.~\ref{fig:WS_coupling_390nm} (right).
The central resonance corresponds to the clock transition frequency $\omega_{eg}$ and  two frequency sidebands located at $\pm\Delta_{g}/2\pi$ arise from weak tunnelling of atoms to neighbouring sites and are completely symmetric with respect to the central resonance.
%This symmetry results in no line pulling of the central resonance. 

The results presented in Figs.~\ref{fig:WS_coupling_390nm} shows that the blue-detuned vertical lattice with depth of $20 E^{\text{rec}}$ has similar level of suppression of the tunelling and of the effects of the atom dynamics as the $10 E^{\text{rec}}$ deep red-detuned optical lattice~\cite{Lemonde2005}. To hold the atoms in a 3D blue-detuned magic lattice, it is required that the energies of the first few lattice states are smaller than the lattice depth. The 20 $E^{\text{rec}}$ trap depth corresponds to around 14 \textmu K and it is sufficiently deep to hold atoms cooled down by using the $^1\text{S}_0$-$^3\text{P}_1$  (red MOT) transition --- typical temperatures achieved in the last cooling stage are of the  order of 1 \textmu K. %\sout{The lattice depth can be higher for all the preparation steps, then we can lower it to remove all atoms except those in the ground oscillation state. The 20 $E^{\text{rec}}$ depth can be applied just for the clock laser interrogation and can be ramped up a bit for state detection.}{

%The blue-detuned lattice with depth of $20 E^{\text{rec}}$ has similar central-to-sideband peak ratio to that of a $10 E^{\text{rec}}$ deep red-detuned optical lattice \cite{Lemonde2005} and therefore providing similar levels of suppression of tunnelling of atoms between different lattice sites.  

\subsubsection{Photoionisation losses due to blue magic lattice}

The previous subsections show that the minimal required optical trap potential depth for an effective operation of an optical strontium clock based on the blue-detuned 390~nm magic wavelength is 
%\textcolor{green}{
on
%} 
the order of $20~E^{\text{rec}}$, assuming vertical probing of the clock transition. {Now we want to estimate the losses due to the presence of the %\sout{390~nm light}
3D blue-detuned lattice potential during a commonly used optical strontium clock operation cycle.}

 %We still consider the 3D lattice potential in the form of Eq.~(\ref{eq:3Dpot}) with $U_{0x}=U_{0y}=U_{0z}=20~E^{\text{rec}}$. 
% \sout{\textcolor{red}{We still consider the 3D lattice potential composed of three independent 1D lattice traps.
%}
%}
%in the form of Eq.~(\textcolor{red}{\ref{eq:3Dpot}})

To characterise the photoionisation losses, we consider  two parts of the clock cycle
%\sout{ (depicted for instance in the right part of Fig.~\ref{fig:timeline_3S1})}: 
 when atoms are being cooled and trapped by the blue MOT and when the atoms are already loaded into the %\sout{red}{3D} 
 optical lattice. We assume that 
 %\sout{the optical}  
 the 3D lattice potential is present during 
 %\textcolor{green}{
 the
 %} 
 whole clock cycle.
 %}
 
 \subsubsection*{Atoms trapped in the blue MOT}
 
The photoionisation losses can lower the total number of the atoms loaded into the blue MOT, which would decrease the number of atoms transferred into the 3D optical lattice trap and  consequently
%\sout{and} 
lower the signal-to-noise ratio of the observed %\sout{detected} 
clock line.
In the blue MOT phase atoms are cooled and trapped through the  $^1$P$_1$ state. 
About $0.002$\% of atoms escape the closed cooling transition $^1$S$_0 \leftrightarrow ^1$P$_1$ and must be repumped. Typically, the repumping through the $^3$S$_1$ state is chosen~\cite{Courtillot05}. However, as the photoionisation of the $^3\text{S}_1$ state by the blue-detuned light at 390~nm is significant, a different repumping scheme, e.g., through the $^3\text{D}$ state~\cite{Mickelson2009,Zhang2020}, is preferable. 
%\textcolor{orange}{\it Here is a weak point - we did not say anything in this subsection about losses during repumping!}

In most of the present realisations of optical lattice clock, the centres of the blue MOT and the lattice trap overlap and the waist of the 1D optical lattice is much smaller than the atomic cloud trapped in the blue MOT. At  the same time, with some notable exceptions~\cite{Baillard07}, the optical lattice is shallow in comparison with the temperature of blue MOT atoms and they can freely pass the optical lattice area.
Therefore, any losses due to photoionisation in the blue MOT due to the blue magic 3D lattice can appear only when atoms are passing the optical lattice region.
%\sout{Atoms passing the optical lattice region experience the average intensity of {$\sim 2.1\times 10^8$~W/m${}^2$} of 390~nm light originating from three independent 1D lattices, each $20~E^{\text{rec}}$ deep.} 

To estimate the losses,  we assume a typical blue MOT condition with maximum total intensity of the trapping beams equal to $I_{461} = 6\times $30~W/m${}^2$ and their detuning from atomic resonance $\Delta = 1.25 \Gamma$,  and use Eq.~(\ref{eq:fraction}) to calculate the relative population of atoms in the $^1$P$_1$ state, $\rho_{^1\text{P}_1}\approx 0.027$.

% Using Fig 9 is misleading here. In OL trap atoms are in the ground state (low intensity), while in MOT they see all intensity

%\textcolor{red}{The photoionisation losses from the   $^1$P$_1$ state for different 3D lattice depths (Fig.~\ref{fig:gamma_vs_U0}) are on the order of $10^4-10^6$~s${}^{-1}$ which is negligible compared to the natural decay rate of $^1$P$_1$ state of $~10^8$~s${}^{-1}$, even for very high lattice depths. {\it So what? Decay rate is not a limiting factor for the number of atoms. We have repumpers!}}}

For the specific case of equal depths of all three blue detuned 1D lattices, and each of the depths equals to $20~E^{\text{rec}}$
%$U_{0x}=U_{0y}=U_{0z}=20~E^{\text{rec}}$ 
considered in the previous subsection, the %\sout{maximum} 
intensity (averaged over time and space) of the 390 nm light experienced by the atoms %\sout{the at the crosing of the} 
passing the optical lattices periodic potentials, 
%region where a MOT and a 3D optical lattice trap overlap - this is not true avarage intesity, it is only true if we would have veeery big waist of lattice beams
is $\sim 2.1\times 10^8$~W/m${}^2$.
The resulting  loss coefficient  (Eq.~(\ref{eq:gamma_P})) is equal to $\gamma_P \approx 2.4 \times 10^5$~s${}^{-1}$.
%rate of losses from the trap in time due to photoionisation is equal to {$~\sim2.4 \times 10^5$} atoms per second.}}
%{\bf Am I right that you assumed Erec potential in each of the lattices directions only and their waists are irrelevant?  Yes, we assume potentials of form $V_{0}*cos^{2}(kx)$} -{\bf and thats is why it is max intesity in the crossing of the beam, with asumption that 3 axises change in time and avarge phase for all 3 beams/Bober}

Assuming that blue MOT lifetime is limited by the collisions with the residual background gas
molecules, the order of magnitude of the loss coefficient $\gamma_{Sr}$ in rate Eqs.~(\ref{eq:rate_full}~-~\ref{eq:sol_loadSr}) in real experimental system can be approximated by the collisional loss rate due to the collisions with H${}_2$ reported in \cite{AbdelHafiz19}, resulting in $\gamma_{Sr}\approx 0.4$~s${}^{-1}$ at the vacuum of $10^{-9}$~mbar. %Adding other channels of losses, e. g. from imperfection of optical repumping, will not increase the $\gamma_{Sr}$ significantly with comparison to $\gamma_P$.%
The loss rate connected with atoms' decay to metastable $^3P_2$ state in the case of operating blue MOT without repumpers is around 35 s$^{-1}$. Therefore, Eq.~(\ref{eq:sol_loadSr}) shows that the blue MOT will be effectively depleted in the region where it overlaps with the  3D optical lattice trap made of three   $20~E^{\text{rec}}$ deep 1D optical lattices. 

%\sout{\textcolor{orange}{Therefore the present of the light at blue-detuned wavelength is not a fundamental limitation for operation of the optical clock for preparing the efficient atomic sample and we can neglect this effect in the future work.}}

To overcome this loss of atoms, one can reduce the lattice intensity or even turn off the lattice beams during the blue MOT phase. %however this can be technically difficult in set-ups using power build-up cavities. 
    Temporary switching off lattice light is technically feasible, e.g. with the power build-up cavity installed inside the vacuum setup on a low expansion glass spacer. The lattice laser can be safely switched back on and relocked during the red MOT phase, which lasts a few tens of ms%\sout{, with simple dedicated relocking electronics}
    . Another possibility is to store cold atoms during the blue MOT phase in the dark $^3P_2$ state \cite{Stellmer2009} or to use cold atomic beam loading directly the red MOT \cite{Chen2019}. 
%Another method {of reducing atomic losses is decreasing the lattice beams' power during the blue MOT phase,} and increasing the power back to trap the atoms in the red MOT phase and to ensure the Lamb-Dicke regime \textcolor{blue}{is restored again} when atoms are interrogated by the clock laser.{

 \subsubsection*{Atoms trapped in the 3D optical lattice}

        After the 
        %\sout{clock} 
        interrogation of the clock transition, the %\sout{usual way of detecting the result of} 
        populations of the ground and %\sout{exciting} 
        excited states %\sout{is} 
        are determined using the optical {repumping} {through}  the $^1$P$_1$ and $^3$S$_1$ states, as discussed before. The effective intensity of the ionising light seen by an atom trapped in the motional ground state of the optical %\sout{trap} 
        lattice is determined by the modulus of its wave function $\psi$, approximated in horizontal and vertical directions by Wannier and Wannier-Stark states, respectively. Due to the %\sout{relative} 
        short lifetimes of the atoms in the  $^3$S$_1$ state, the %\sout{spacial}
        spatial distribution of the atoms is determined by their preceding, metastable  $^3$P$_0$ state (the short lifetime prevents changing the shape of the distribution in the new potential corresponding to the $^3$S$_1$ state polarisability before they decay to the lower states, which are not susceptible to photoionisation). 
        The ${}^{88}$Sr atoms have the same polarisabilities 
%\sout{at} 
in the $^3$P$_0$ and $^1$S$_0$ states {at} %\textcolor{blue}{for the blue-detuned magic wavelength } 
{the optical trapping magic wavelength} (at 389.889~nm these polarisabilities are equal to 459~a.u. as calculated using the data in Ref.~\cite{Zhou10}), therefore in our calculations, we have used numerically calculated %\sout{can use} 
Wannier and Wannier-Stark states %
%\sout{calculated numerically} 
for a given potential scaled in the units of $E^{\text{rec}}$. 

The effective intensity from 1D lattice along $\zeta$ axis is calculated by

\begin{equation}
    I_{\text{eff},\zeta}=\int |\psi(\zeta)|^2 I(\zeta) d\zeta, \text{ with } \zeta = x,y,z,
    \label{eq:Ieff}
\end{equation}

\noindent
w
here $I(\zeta)=I_{0\zeta}\sin^2(k_\zeta  \zeta)$ is the standing wave intensity distribution and $I_{0\zeta}$  is the maximum intensity of each of 
1D traps 
%\textcolor{red}{1D traps} 
in $\zeta$ direction. It should be pointed out that the effective intensity $I_{\text{eff},\zeta}$ is calculated differently than the average intensity $I_{P}$ from Eq.~\ref{eq:avr_intensity}. $I_{P}$ is the average intensity of light  from a single non-reflected  photoionization beam with Gaussian intensity distribution, whereas the $I_{\text{eff},\zeta}$ is the effective intensity which induces photoionization of atoms in the ground Wannier or Wannier-Stark state of the blue-detuned optical lattice periodic potential.
%\sout{\textcolor{red}{\it Here, or  earlier, the relation between $I_{0\zeta}$ and $U_{0\zeta}$.}}

\begin{figure}[hbt]
\centering
\includegraphics[width=0.5\columnwidth]{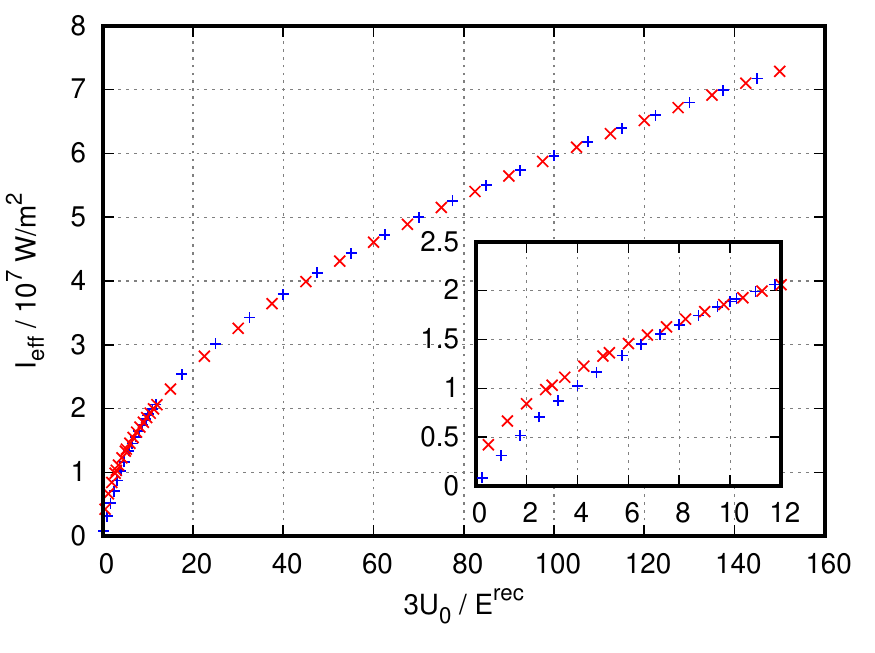}%in the form of Eq.~(\textcolor{red}{\ref{eq:3Dpot}})

\caption{The total effective intensity $I_\text{eff}$ experienced by the atoms trapped in the motional ground state of a 3D blue-detuned magic wavelength optical lattice. $I_\text{eff}$ values were calculated for Wannier-Stark (Wannier) states for the vertical (horizontal) directions in the optical lattice (blue crosses) and Gaussian %\sout{harmonic oscillator} 
states (red crosses) in a harmonic potential approximation.%\sout{, both according to Eq.~\ref{eq:Ieff}}.% and fitted with a root-square function (black solid line). 
% \sout{lattice experienced by atoms in the ground state on the trapping potential.}
\label{fig:intensity_percentage} }
\end{figure}

%\textcolor{red}{Sentence can not be started with shortened word like fig. or we have to change the structure of the sentence or we have to write Figure (full word)}
Figure~\ref{fig:intensity_percentage} depicts 
%\textcolor{red}{the dependence of the total effective intensity in 3D lattice $I_{\text{eff}} = \sum_{\zeta} I_{\text{eff},\zeta}$
the dependence of the total effective intensity in 3D lattice $I_{\text{eff}} = \sum_{\zeta} I_{\text{eff},\zeta}$ on the total amplitude of the potential $3U_{0}$.
The blue crosses depict the values numerically calculated for Wannier-Stark (Wannier) states for the vertical (horizontal) directions in an
optical lattice. In general, for deep enough optical lattice traps the calculations can be greatly simplified by approximating the system in Eq.~(\ref{eq:Ieff})  by a harmonic potential and its Gaussian ground states, which yields a square-root dependence of the effective intensity on the lattice potential (red crosses).
%\textcolor{blue}{(blue crosses) \textbf{something about why we use the HO approx also?}}.% with a fit of the square-root function (solid black line).}
%\sout{the relative intensity of blue-detuned lattice light $I_{\text{eff}}/(4I_{lat})$ seen by atoms in the ground state.}

%\begin{figure}[hbt]
%\centering
%\includegraphics[width=0.6\columnwidth]{gamma.eps}
%\caption{\textcolor{blue}{To be removed }\textcolor{red}{Calculated photoionization rates for $^1$P$_1$ (blue) and $^3$S$_1$ (red) states for different 3D lattice depths for 390~nm magic wavelength lattice {\it This is for atoms trapped in the ground state of the lattice, or for atoms seeing the total intensity???} }}
%\label{fig:gamma_vs_U0} 
%\end{figure}

%Using Eq. \ref{eq:gamma_P} and Eq.\ref{eq:Ieff}, we calculate the photoionization losses for $^3$S$_1$ state shown in Fig \ref{fig:gamma_vs_U0}. The photoionization losses are of the order of $10^3$ atoms per second.
%\textcolor{red}

In the considered lattice consisting of three independent 1D lattices, each $20~E^{\text{rec}}$ deep, the $I_{\text{eff}} \sim 4.6\times 10^7$~W/m${}^2$. To calculate the rate of losses for $^3$S$_1$ state, we replace in Eq. \ref{eq:gamma_P} the average intensity $I_{P}$ with the effective intensity $I_{\text{eff}}$. The resulting rate of losses of atoms in the $^3$S$_1$ state  due to photoionisation is %\textcolor{green}{\sout{less than}} 
$1.26\times 10^8$~s$^{-1}$, which  
%\textcolor{red}{ 
is of the order of 
the $^3$S$_1$ state decay rate due to
%} 
the natural lifetime.
Such a large loss rate significantly limits the applicability of the blue magic lattice with a $^3\text{S}_1$-based repumping scheme. 
A possible solution is to employ an alternative repumping scheme through the $^3\text{D}_{1,2}$ states {(see Fig.~\ref{fig:levels})}. {As these $^3\text{D}$ states lie below the photoionisation threshold for blue-detuned optical lattice, they are not affected by the photoionisation induced losses from ~390 nm wavelength light. However, the high IR wavelengths of 2.6~\textmu m for the $^{3}$P$_{0}$-$^{3}$D$_{1}$ and 3~\textmu m for the $^{3}$P$_{2}$-$^{3}$D$_{2}$ transition make the use of these transitions experimentally challenging as the lasers at these wavelengths are often not readily available. Additionally, the longer lifetimes of $\mathrm{\tau_{^{3}D_{1}}=2.18(1)}$~\textmu s for $^{3}$D$_{1}$\cite{Nicholson2015} and $\mathrm{\tau_{^{3}D_{2}}=12.7}$~\textmu s for $^{3}$D$_{2}$ state {(the value deduced from the energy diagram in \cite{Mickelson2009})} would require longer repumping times during the detection of the clock transition.}\\
%MW and MZ: tau for 3d2 from Mickelson isnt measured, it's simply in their energy diagram figure. I couldnt find the actual lifetime for 3D2.}   \sout{state%~\cite{Mickelson2009,Zhang2020}, 
%which is not ionised by the blue magic light as it lies below the ionisation threshold.} 

%}

\subsubsection{AC Stark shift from independent 1D lattices}
To prevent interference between independent 1D lattices,
small frequency detunings must exist between the light of each of the 1D lattices. While the detuning from the magic wavelength of the 1D lattice optical clock would add a considerable light shift to the clock frequency, in the 3D lattice the effective light shift can be still controlled at the required level. 
%One can assume that in the clock laser interrogation axis the 1D optical lattice can be precisely tuned to the magic wavelength. 
One can assume that the 1D optical lattice in the clock laser interrogation axis can be precisely tuned to the magic wavelength. 
With blue detuned optical lattice, all higher-order effects will be suppressed as atoms are trapped close to the intensity minimum. %\sout{As was calculated} 
According to our calculations, a trapping depth of 20~$E^\text{rec}$ is enough for clock operations in the vertical direction. Tunnelling in horizontal directions do not have to be suppressed that well, and we assume individual trap depth to be 20~$E^\text{rec}$ as well.  With the vertical lattice tuned to the magic wavelength, the required minimal detuning of two other individual lattice beams is determined by the atomic oscillation period and thus by the trap frequency. For a trapping depth of 20~$E^\text{rec}$, the trap frequency is around 135~kHz. With detuning %\sout{of} 
around a few times the trap frequency, we can assume that potential can trap atoms efficiently. Both horizontal lattice beams can be detuned to opposite sides of magic wavelength, which in case of identical intensity and almost perfectly linear light shift scaling around magic wavelength \cite{Takamoto2009} should cancel out individual light shifts. If we assume detuning of plus and minus 500~kHz of each  horizontal lattice beam and a relative intensity difference of 3\%, the induced effective light shift from detuning from both horizontal lattice beams can be controlled to the level of around $1 \times 10^{-19}$. For both horizontal lattices detuned by 1~MHz and an %\sout{with} 
intensity mismatch of 5\%, the induced effective light shift is still around $5 \times 10^{-19}$. For this estimation, we use %\sout{calculated}
the dependence of the intensity seen by atoms calculated in subsection 3.2.3.

\section{Conclusion}
In conclusion, we have determined the values of the $^{88}$Sr photoionisation cross section at blue-detuned magic wavelength 389.889~nm 
%\textcolor{red}{
to be 2.20(50)$\times$10$^{-20}$~m$^2$ and %4.5(4.5)$\times$10$^{-23}$~m$^2$, 
$1.38(66)\times$10$^{-18}$~m$^2$,
for the excited states $^1$P$_1$ and $^3$S$_1$, respectively.
%} 
Additionally, we have measured the photoionisation cross section for the $^1$P$_1$ state in a range from 378.4~nm to 407~nm and determined
the position and the peak value of the autoionisation resonance (4d$^2$+5p$^2$) $^1\text{D}_2$ 
to be 405.196(44)~nm and 5.20(94)$\times$10$^{-19}$~m$^2$, respectively. 
These results are consistent with the results previously reported~\cite{Mende1995,Sami-ul-Haq2006}, which, unlike ours, did not involve cold atomic systems.

To examine the potential feasibility of the blue-detuned %\textclor{red}{-detuned} 
magic  wavelength strontium optical lattice trap, we have estimated photoionisation-induced atomic losses  %\textcolor{red}{
in a three-dimensional optical lattice trap operating at the minimal intensity required to %\sout{prevent the Doppler and recoil shifts of the clock transition} 
fulfil the Lamb-Dicke regime~\cite{Dicke1953,Wineland1979} for the clock transition. For a 3D lattice with lattice depths of $20 E^\text{rec}$ we found the rate of photoionization losses for $^1$P$_1$ and $^3$S$_1$ to be $\gamma_P \approx 2.4 \times 10^5$~s${}^{-1}$ and $1.26\times 10^8$ s$^{-1}$, respectively, and we have compared these losses with other channels of atomic losses during the standard operation of an optical lattice strontium clock. We also make several suggestions on %\sout{omit}
mitigating the photoionisation losses for $^1$P$_1$ and $^3\text{S}_1$ states, thus ensuring that neither of these loss channels %\sout{are}
is a critical defect for a blue-detuned lattice clock. {In particular, the large photoionisation loss rate in the blue-detuned optical lattice makes the use of $^{3}$S$_{1}$ state in the optical clock cycle unfeasible and would instead require the use of the less commonly $^{3}$D$_{1}$ state which, while feasible, adds additional experimental difficulties due to the high IR wavelengths of the relevant ${}^{3}$P${}_{0}$-${}^{3}$D${}_{1}$ and ${}^{3}$P${}_{2}$-${}^{3}$D${}_{2}$ transitions. }
%
%\sout{for typical experimental conditions. We noted very little influence of the ionising effect on the performance of the Sr optical lattice clocks.}\textcolor{red}{\it This is not true - blue MT is highly disturbed.}
%

{Interestingly, the non-destructive measurements of the clock transition probability(e.g. \cite{Vallet17,Hobson19}), assuming a different way of repumping,  still seem compatible with the blue-detuned trap. The non-destructive measurements that utilise 461~ńm light assume that the excited level is not populated thanks to detuning and low power of the local oscillator and probe beams (down to $\sim$30~uW at the waist of $\sim$75~um and detuning of $\sim$2~GHz in the newest experiments, that corresponds to the relative population of atoms in the $^1$P$_1$ state, $\rho_{^1\text{P}_1}\approx 0.00025$), because any atom in the excited $^1$P$_1$ state in the optical lattice is lost from the relatively shallow optical trap.}

{Moreover, the schemes of a continuous superradiant optical active clocks that were proposed in ~\cite{Kazakov13} and related works  
that use a blue-detuned magic wavelength to realize the Lamb-Dicke regime are not using either the first-stage cooling or excitation readout in the presence of the blue-detuned optical lattice, therefore their feasibility is not affected by high photoionisation cross sections. The losses can, however, impact newer proposals, that would improve the proposed scheme by including repumping of superradianting atoms by external incoherent pumping~\cite{Hotter19,Bychek21}}.

\section*{Funding}

H2020 Future and Emerging Technologies (No 820404, iqClock project); The European Metrology
Programme for Innovation and Research (EMPIR) Programme cofinanced by the Participating States and from the
European Union’s Horizon 2020 Research and Innovation Programme (EMPIR 17FUN03 USOQS); Fundacja na rzecz Nauki Polskiej (TEAM/2017-4/42); Narodowe Centrum Nauki (2017/25/B/ST2/00429, 2017/25/Z/ST2/03021).
% The “A next-generation worldwide quantum sensor network with optical atomic clocks” project (TEAM/2017-4/42) is carried out within the TEAM IV Programme of the Foundation for Polish Science co-financed by the European Union under the European Regional Development Fund. This project has received funding from the EMPIR Programme co-financed by the Participating States and from the European Union’s Horizon 2020 Research and Innovation Programme (EMPIR 17FUN03 USOQS). This project has received funding from the European Union’s Horizon 2020 Research and Innovation Programme No 820404, (iqClock project).
%The “A next-generation worldwide quantum sensor network with optical atomic clocks” project is carried out within the TEAM IV Programme of the Foundation for Polish Science cofinanced by the European Union under the European Regional Development Fund. Support has been received from the project EMPIR 17FUN03 USOQS. This project has received funding from the EMPIR Programme cofinanced by the Participating States and from the European Union’s Horizon 2020 Research and Innovation Programme. 
% M.~Witkowski,  D.~Kovačić, and V.~Singh are supported by the National Science Centre, Poland, Project no. 2017/25/B/ST2/00429. M.~Bober and A.~Tonoyan are supported by the National Science Centre, Poland, under QuantEra programme no. 2017/25/Z/ST2/03021. 
\section*{Acknowledgements}

The measurements were performed at the National Laboratory FAMO (KL FAMO) in Toruń, Poland, and were supported by a subsidy from the Polish Ministry of Science and Higher Education.

% \section*{Acknowledgments}

\section*{Disclosures}

The authors declare no conflicts of interest.

\section*{Data availability}

Data underlying the results presented in this paper are available in the open data repository~\cite{repod}.

\bibliography{main}

%%%%%%%%%% If preparing manually:
%  \begin{thebibliography}{1}
% \newcommand{\enquote}[1]{``#1''}

% \bibitem{Zhang:14}
% Y.~Zhang, S.~Qiao, L.~Sun, Q.~W. Shi, W.~Huang, L.~Li, and Z.~Yang,
%   \enquote{Photoinduced active terahertz metamaterials with nanostructured
%   vanadium dioxide film deposited by sol-gel method,}
%   {\protect\JournalTitle{Optics Express}} \textbf{22}, 11070--11078 (2014).

% \bibitem{OSA}
% {Optical Society}, \enquote{{OSA Publishing},}
%   \url{http://www.osapublishing.org}.

% \bibitem{FORSTER2007}
% P.~Forster, V.~Ramaswamy, P.~Artaxo, T.~Bernsten, R.~Betts, D.~Fahey,
%   J.~Haywood, J.~Lean, D.~Lowe, G.~Myhre, J.~Nganga, R.~Prinn, G.~Raga,
%   M.~Schulz, and R.~V. Dorland, \enquote{Changes in atmospheric consituents and
%   in radiative forcing,} in \enquote{Climate Change 2007: The Physical Science
%   Basis. Contribution of Working Group 1 to the Fourth assesment report of
%   Intergovernmental Panel on Climate Change,}  S.~Solomon, D.~Qin, M.~Manning,
%   Z.~Chen, M.~Marquis, K.~B. Averyt, M.~Tignor, and H.~L. Miler, eds.
%   (Cambridge University Press, 2007).

%  \end{thebibliography}

\end{document}